\documentclass{elsart}

 \usepackage{graphicx}

\usepackage{amssymb}

\begin{document}

\begin{frontmatter}



\title{Kinetics of helium bubble formation in nuclear materials }

\author[UC3M]{ L. L. Bonilla \corauthref{cor}},
\corauth[cor]{Corresponding author.}
\ead{bonilla@ing.uc3m.es}
\author[UCM]{A. Carpio},
\ead{ana$_{-}$carpio@mat.ucm.es}
\author[UCB]{J. C. Neu}, 
\ead{neu@math.berkeley.edu}
\author[LLNL]{ W. G. Wolfer},
\ead{wolfer1@llnl.gov }

\address[UC3M]{Grupo de Modelizaci\'on, Simulaci\'on Num\'erica y Matem\'atica
Industrial, 
Universidad Carlos III de Madrid, Avenida de la Universidad 30, 28911 Legan{\'e}s, Spain}
\address[UCM]{Departamento de Matem\'{a}tica Aplicada, Universidad
Complutense de Madrid, 28040 Madrid, Spain}
\address[UCB]{Department of Mathematics, Universidad de California at Berkeley, 
Berkeley, CA 94720; USA}
\address[LLNL]{Lawrence Livermore National Laboratory, Livermore, CA 94550; USA}
\date{ 17 July 2006  }

\begin{abstract}
 The formation and growth of helium bubbles due to self-irradiation in plutonium has been 
 modelled by a discrete kinetic equations for the number densities of bubbles having $k$ 
 atoms. Analysis of these equations shows that the bubble size distribution function can be 
 approximated by a composite of: (i) the solution of partial differential equations 
 describing the continuum limit of the theory but corrected to take into account the effects of 
 discreteness, and (ii) a local expansion about the advancing leading edge of the distribution 
 function in size space. Both approximations contribute to the memory term in a close 
 integrodifferential equation for the monomer concentration of single helium atoms. 
 The present boundary layer theory for discrete equations is compared to the numerical 
 solution of the full kinetic model and to previous approximation of Schaldach and Wolfer 
 involving a truncated system of moment equations. 
\end{abstract}

\end{frontmatter}

\section{Introduction}
There are simple kinetic models of irreversible aggregation, in which a cluster with $k$
monomers grows by absorbing one monomer but it cannot decrease in size by shedding
part of its mass. An interesting example is the formation and growth of helium bubbles
in plutonium alloys as a consequence of alpha decay due to self-irradiation 
\cite{SW04,SWZW05}. As an alloy ages, there is an initial transient stage during which 
self-irradiation produces dislocation loops that tend to saturate within approximately two 
years. The alpha particles created during irradiation become helium atoms. These atoms 
come to rest at unfilled vacancies generated during their slowing-down process, before they 
are captured at existing helium bubbles. A helium atom diffuses through the lattice until it
finds another helium atom thereby forming a stable dimer or until it finds a helium bubble (a 
stable cluster with $k$ atoms or, in short, a $k$-cluster), which absorbes it. Helium bubbles are 
attached to lattice defects, do not move and do not shed helium atoms because the binding 
energies of helium to any cluster are extremely high \cite{SWZW05}. The following 
kinetic model based on these observations has been proposed by Schaldach and Wolfer 
\cite{SW04}:
\begin{eqnarray}
\dot{\rho}_{k} =4\pi D\, \rho_{1}\, a_{k-1}\rho_{k-1}- 4\pi D\, \rho_{1}\, 
a_{k}\rho_k,\quad  k\geq 3,    \label{e1}\\ 
\dot{\rho}_{2} =8\pi D\, \rho_{1}^2\, a_{1} - 4\pi D\,\rho_{1}\, a_{2}\rho_2,   \label{e2}\\
\rho_{1} + \sum_{k=2}^\infty k \rho_k =  \int_{0}^{\tilde{t}} g(t')\, dt'.  
\label{e3}
\end{eqnarray}
Here $\dot{\rho}_{k} =d\rho_{k}/d\tilde{t}$, $\rho_k$ is the number density of $k$ 
clusters having effective radii $a_{k}$ (when the center of a monomer comes within distance 
$a_{k}$ of the cluster center, it is absorbed), $\rho_{1}$ is the number of monomers per 
unit volume, $D$ is the diffusion coefficient and $g(\tilde{t})$ is the number of monomers 
created per unit volume and per unit time. Eq.\ (\ref{e3}) means that the total number of 
monomers per unit volume, whether they are in solution or forming part of a $k$ cluster, 
should equal the time integral of $g(\tilde{t})$. In Equations (\ref{e1}) and (\ref{e2}), 
$k$ clusters grow by adding one monomer with a rate $4\pi D\rho_{1}a_{k-1}$ (for $k>2$) 
to a $k-1$ cluster, and they do not decay. The mean absorption rate of a monomer by an 
immobile $k$ cluster (with $k>1$) is $4\pi D\rho_{1} a_{k}$, whereas the rate of creation 
of an immobile dimer by the collision of two mobile monomers is twice this quantity, $8\pi 
D\rho_{1}a_{1}$. The absorption rate can be calculated immediately from the concentration 
field outside a $k$ cluster (i.e., the number of monomers per unit volume in $\tilde{r}>
a_{k}$), $\rho(\tilde{r},\tilde{t})$ assuming that the {\it mean-field approximation} 
holds. The process of diffusion-limited adiabatic growth of clusters is random: diffusion of 
monomers to a given cluster is randomly affected by the spatial distribution of the other
clusters in its vicinity. If $a_{k}^3\rho_{k}\ll 1$ (diluted clusters), we may assume that
the concentration field outside a cluster goes to the average monomer density in solution,
$\rho_{1}(\tilde{t})$, which ignores fluctuations in the local environment and constitutes 
the mean-field approximation. Then $\rho(\tilde{r},\tilde{t})$ solves the Laplace 
equation outside $\tilde{r}=a_{k}$, with $\rho(a_{k},\tilde{t})=0$, and it takes on the 
value $\rho_{1}(\tilde{t})$ as $\tilde{r}\to\infty$. Thus
$$ \rho(\tilde{r},\tilde{t})= \rho_{1}(\tilde{t})\,\left(1-{a_{k}\over\tilde{r}}
\right),
$$ 
for $\tilde{r}>a_{k}$. The mean absorption rate of monomers by the $k$ cluster is
\begin{eqnarray}
4\pi a_{k}^2 D{\partial\rho\over\partial\tilde{r}}(a_{k},\tilde{t}) = 4\pi D 
\rho_{1} a_{k}.  \label{e4}
\end{eqnarray}
The simplest model for the effective radius of a $k$ cluster is based on packing of 
non-overlapping particles:
\begin{eqnarray}
a_{k} = a_{1}\, k^{1/3}.    \label{e5}
\end{eqnarray}

The mean-field model (\ref{e1}) - (\ref{e5}) has obvious limitations: the kinetic equations
are deterministic due to the mean-field approximation while the creation of helium atoms
and diffusion of monomers are random processes, and helium bubbles have limitations to their
growth beyond a certain size. These effects will not be addressed in the present paper.  
It is interesting to observe that a related kinetic system was proposed and solved in 1914
by McKendrick as a model of leucocyte phagocytosis \cite{mck14}. In McKendrick's model,
$\rho_{k}$ is the density of leucocytes which have ingested $k$ bacteria, and its rate 
equation is (\ref{e1}) for $k\geq 0$, with a known function of time $\rho_{1}(\tilde{t})
>0$ and $\rho_{-1}\equiv 0$. McKendrick's solution method involved solving the equation 
for $\rho_{0}$ in terms of $\int \rho_{1}\, d\tilde{t}$ and solving recursively all 
other equations for $\rho_{k}$ as functions of $\rho_{0}$. His method cannot be used to 
solve the system (\ref{e1}) - (\ref{e3}), but an useful closure of this infinite system to 
only three differential equations was introduced in \cite{SW04}, and compared to 
experiments, \cite{SW04,SWZW05}. 

In this paper, we shall study the solution of the system (\ref{e1}) - (\ref{e3}) starting 
from an initial condition corresponding to the absence of helium bubbles, i.e.,
\begin{eqnarray}
\rho_{k}(0)=0,\, \mbox{for}\, k\geq 1. \label{e6}
\end{eqnarray}
We shall consider the case of a constant production rate of helium atoms, $g(\tilde{t})=g\, 
\tilde{t}$. Once $\rho_{1}(\tilde{t})$ as been found, the total density of bubbles can 
be obtained by integrating the equation:
\begin{eqnarray}
\frac{d}{d\tilde{t}}\sum_{k=2}^\infty \rho_k = 8\pi D a_{1} \rho_{1}^{2}.  
\label{e7}
\end{eqnarray}
Eq.\ (\ref{e7}) is immediately obtained by adding (\ref{e2}) and (\ref{e1}) (for all $k
\geq 3$). 

We have found that, after a short transient, the size distribution function becomes a smooth
function of $k$ and it is a functional of the monomer concentration. In turn, the monomer 
concentration satisfies an integrodifferential equation. The size distribution function can
be approximated by matched asymptotic expansions. Except for large $k$, its 
form is close to the solution of the hyperbolic equation resulting from replacing derivatives 
instead of differences in (\ref{e1}), but including corrections due to discreteness. This
outer approximation breaks down at large $k$: it has an unrealistic large peak at a maximum
value of $k$ after which the distribution function abruptly falls to zero. If our governing
equations were partial differential equations instead of differential-difference equations, 
we would say that a boundary layer should be inserted to remedy this imperfection. However,
the concept of boundary layer is not straightforward for discrete equations. How do we 
insert a boundary layer approximation to difference equations? The answer to this question
lies in the wave front expansion of the Becker-D\"oring equations some of us introduced in
Ref.~\cite{NBC05}. This theory yields one equation for the leading edge of the size 
distribution function and one equation for the distribution function itself near its edge. 
A solution of the latter written in similarity variables is then matched to the outer 
approximation and its contribution to the integral terms of the equation for the monomer 
concentration is calculated. These two steps were not needed for the nucleation calculation in 
Ref.~\cite{NBC05} because there the outer approximation was a constant, size-independent
profile and the integral condition was identically satisfied during the nucleation transient. 
The present overall theory including outer and boundary layer approximations describes 
quite well the observed numerical simulations of the discrete system of equations. 

The rest of the paper is as follows. The nondimensional equations of the model are derived in 
Section \ref{nodim}. The outer approximation to the size distribution function and the
monomer concentration is described in Section \ref{similarity}. The boundary layer analysis
is given in Section \ref{front}. Sec.~\ref{sec:wolfer} compares the present theory to a 
simple system of three differential equations for the helium density, the bubble density and
the monomer concentration obtained by closing the moment equations for the size
distribution according to a simple ansatz introduced in \cite{SW04}. The last section 
summarizes our conclusions, and technical details are relegated to Appendices.

\section{Nondimensionalization of the kinetic equations}
\label{nodim}
Let $[t]$, $[k]$, $[c]$, $[\rho]$ be typical units of time, cluster size, monomer 
concentration and cluster density, respectively. Notice that we use $[c]$ instead of the more
cumbersome notation $[\rho_{1}]$ in order to avoid confusion between the units of monomer
concentration and cluster density. Equations (\ref{e1}), (\ref{e7}) and 
(\ref{e3}) provide the following dominant balances between these scaling units:
\begin{eqnarray}
&&\frac{[\rho]}{[\tilde{t}]} =4\pi D\, [c]\, a_{1}\frac{[k]^{1/3} 
[\rho]}{[k]},\label{e8}\\ 
&& \frac{[\rho]\, [k]}{[\tilde{t}]} = 4\pi D\, [c]^2\, a_{1},   
\label{e9}\\
&& [k]^2 [\rho] =  g\, [\tilde{t}].  \label{e10}
\end{eqnarray}
To derive these equations, we have assumed that there is a continuum limit with $k\gg 1$
and that the number density of particles in clusters, $\sum_{k=2}^\infty k\, \rho_{k}$,
is much larger than the monomer concentration $\rho_{1}$. Note that we have three 
relations (\ref{e8}) - (\ref{e10}) for four unknowns $[\tilde{t}]$, $[k]$, $[c]$, $[\rho]$. 
In the absence of another relation (for example, a different creation rate $g(\tilde{t})
=G(\tilde{t}/\tau)$, which fixes the time scale $[\tilde{t}]=\tau$), we can express three 
of these scaling units in terms of the fourth. We find
\begin{eqnarray}
[\tilde{t}] = \frac{[k]^{7/6}}{\sqrt{4\pi D a_{1}g}}, \quad  [c]= 
\sqrt{{g\over 4\pi Da_{1}}}\, [k]^{-1/2}, \quad [\rho] =  \sqrt{{g\over 4\pi D
a_{1}}}\, [k]^{-5/6}.  \label{e11}
\end{eqnarray}

We shall now define nondimensional units as
\begin{eqnarray}
t={\tilde{t}\over [\tilde{t}]} \equiv \kappa^{-7/6}\,\tilde{t}\sqrt{4\pi D 
a_{1}g}, \nonumber\\
c={\rho_{1}\over  [c]} \equiv \kappa^{1/2}\rho_{1}\,\sqrt{{4\pi D
a_{1}\over g}}, \nonumber\\
r_{k}= {\rho_{k}\over [\rho]}\equiv  \kappa^{5/6}\rho_{k}\sqrt{{4\pi Da_{1}
\over g}},  \label{e12}
\end{eqnarray}
in which we have set $[k]=\kappa$ (any positive number). Equations (\ref{e1}) - 
(\ref{e3}) and (\ref{e7}) become
\begin{eqnarray}
&& {dr_{k}\over dt} = c\, \kappa^{2/3}\, [(k-1)^{1/3}r_{k-1}- k^{1/3} r_k],
\quad  k\geq 3,    \label{e13}\\ 
&& {dr_{2}\over dt}= 2 \kappa\, c^2 - \kappa^{2/3}2^{1/3}c\, r_2,
\label{e14}\\
&& \kappa^{-5/3} c +\kappa^{-2}\sum_{k=2}^\infty k r_k =  t,  \label{e15}\\
&& {d\over dt}\sum_{k=2}^\infty r_{k} = 2 \kappa\, c^2. \label{e16}
\end{eqnarray}

Our nondimensional system of kinetic equations comprises (\ref{e13}) - (\ref{e15}) with 
$\kappa=1$ and initial conditions $r_{k}(0)=0$, $c(0)=0$. (\ref{e16}) is a consequence 
of the previous equations or it can be used instead of (\ref{e14}). Defining an adaptive
time $s=\int_{0}^{t} c(t')\, dt'$, we can rewrite our system in the more 
convenient form:
\begin{eqnarray}
&& {dr_{k}\over ds} = (k-1)^{1/3}r_{k-1}- k^{1/3} r_k,  \quad  k\geq 3,  \label{e17}\\ 
&&{dr_{2}\over ds}= 2 c - 2^{1/3}r_{2},  \label{e18}\\
&& c + \sum_{k=2}^\infty k r_k =  t,  \label{e19}\\
&& {ds\over dt} = c, \label{e20}
\end{eqnarray}
with initial conditions $c(0)=r_{k}(0)=0$ ($k\geq 2$) and $t(0)=0$. In order to
integrate this system of equations, it is convenient to time differentiate (\ref{e19}) and use 
(\ref{e20}). The result is 
\begin{eqnarray}
c\, {dc\over ds}+ 4\, c^2 + c\, M_{1/3} = 1,  \label{e21}
\end{eqnarray}
in which we have defined the moments of the size distribution function $r_{k}(s)$:
\begin{eqnarray}
M_{\mu}(s)= \sum_{k=2}^\infty k^\mu r_{k}.  \label{e22}
\end{eqnarray}
$M_{0}$ and $M_{1}$ are the number densities of bubbles and of helium, respectively. They 
satisfy:
\begin{eqnarray}
{dM_{0}\over ds} &=& 2 c,  \label{e23}\\
{dM_{1}\over ds} &=& 4c + M_{1/3},  \label{e24}
\end{eqnarray}
with zero initial conditions. 

The kinetic equations describing formation of helium bubbles are therefore (\ref{e17}),
(\ref{e18}), (\ref{e20}), (\ref{e21}) and (\ref{e22}) with zero initial conditions. These
equations yield the monomer concentration $c$ and the size distribution function $r_{k}$. 
The number densities of bubbles and of helium are given by (\ref{e23}) and (\ref{e24}).

\section{Outer solution and relation to the continuum limit equations}
\label{similarity}
\subsection{Continuum limit and similarity solution}
A first attempt at approximating the model equations consists of taking the continuum limit 
of (\ref{e17}) - (\ref{e20}). In fact, assume that 
\begin{eqnarray}
r_{k}(s)=r(k,s), \label{e38} 
\end{eqnarray}
and Taylor expand $r$ in (\ref{e17}) up to first order terms. The result is
\begin{eqnarray}
&& {\partial r\over \partial s} + {\partial\over\partial k}(k^{1/3} r) = 0, 
\label{e39}\\ 
&& \int_{0}^{\infty} k\, r\, dk =  t.   \label{e40} 
\end{eqnarray}
We have ignored the monomer concentration in (\ref{e40}). Integrating (\ref{e39}) over 
$k>0$, we obtain $$ {d\over ds}\int_{0}^\infty r\, dk = \lim_{k\to 0}(k^{1/3} r), $$
and (\ref{e23}) implies the following signaling condition:
\begin{eqnarray}
\lim_{k\to 0}(k^{1/3} r) = 2 c.   \label{e41}
\end{eqnarray}

The method of characteristics provides the following solution to (\ref{e39}) and 
(\ref{e41}) with initial condition $c(0)=0$:
\begin{eqnarray}
r(k,s) = 2 k^{-1/3}\, c\left(s-a(k)\right)\, \theta\left(s-a(k)\right), \label{e42}\\
a(k)= {3\over 2}\, k^{2/3},  \label{e42a}
\end{eqnarray}
in which $\theta(x)=1$ if $x>0$ and $\theta(x)=0$ if $x<0$ is the Heaviside unit step
function. After a change of variable, the integral condition (\ref{e40}) becomes
\begin{eqnarray}
2\, \left({2\over 3}\right)^{3/2}\int_{0}^{s} (s-s')^{3/2}\, c(s')\, ds' = t, 
\label{e43}
\end{eqnarray}
or, equivalently,
\begin{eqnarray}
2\, \left({2\over 3}\right)^{1/2} c(s)\,\int_{0}^{s} (s-s')^{1/2}\, c(s')\, ds' = 1. 
\label{e44}
\end{eqnarray}

The problem (\ref{e39}) - (\ref{e41}) and (\ref{e43}) has a similarity solution whose 
role will be discussed later. In fact, note that (\ref{e39}) - (\ref{e41}) are invariant under the scaling 
transformation $r\to\kappa^{-5/6} r$, $c\to\kappa^{-1/2} c$, $s\to\kappa^{2/3} s$, 
$t\to\kappa^{7/6}t$, $k\to\kappa k$, which is suggested by the nondimensionalization 
(\ref{e12}). Therefore the combinations $\chi=k\, s^{-3/2}$ and $\rho=s^{5/4}r$ are 
invariant under the previous scaling transformation. In terms of $\rho=\rho(\chi)$, 
(\ref{e39}) becomes
\begin{eqnarray}
&& {d\rho\over ds} = {\rho\over 3\chi}\, {{15\over 4}\,\chi^{2/3} - 1\over 
{3\over 2}\,\chi^{2/3}},    \label{e45}\\ 
&& \rho= s^{5/4}\, r,\quad \chi = k\, s^{-3/2}.  \label{e46}
\end{eqnarray}
The solution of (\ref{e45}) is 
\begin{eqnarray}
r(\chi,s) = {R_{0}s^{-5/4}\chi^{-1/3}\over \left(1-{3\over 2}\,\chi^{2/3}
\right)_{+}^{3/4}}.        \label{e48}
\end{eqnarray}
Here $R_{0}$ is a positive constant and $f(x)_{+}= f(x)\,\theta(x)$ for any function 
$f(x)$. This similarity solution has integrable singularities at $\chi=0$ and at $\chi=
\chi_{0}\equiv (2/3)^{3/2}$ (corresponding to the position of the local maximum of
$r_{k}$ for large sizes, $k_{M}(s)$, which coincides with the maximum cluster size in this
case). Now the signaling condition (\ref{e41}) yields the monomer concentration:
\begin{eqnarray}
c(s) = {1\over 2}\, R_{0}\, s^{-3/4},       \label{e49}
\end{eqnarray}
and (\ref{e20}) together with $s(0)=0$ yield the relation between $s$ and the time:
\begin{eqnarray}
s = \left({7\over 8}\, R_{0}\, t\right)^{4/7}.        \label{e50} 
\end{eqnarray}
The constant $R_{0}$ is found by inserting (\ref{e49}) in (\ref{e44}), which yields 
$R_{0}^2\sqrt{2/3} B(3/2,1/4)=1$, from which,
\begin{eqnarray}
R_{0}= {(27\pi)^{1/4}\over\Gamma(1/4)}\approx 0.837042. \label{e51} 
\end{eqnarray}
\begin{figure}
\begin{center}
\includegraphics[width=10cm]{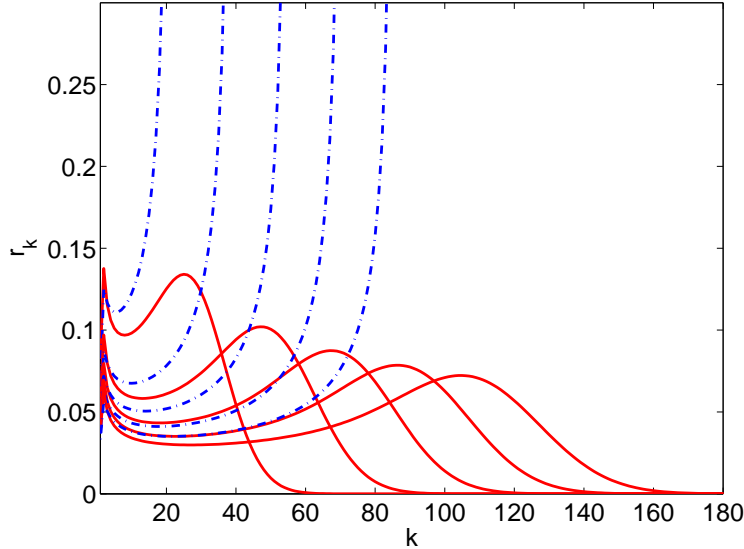}
\vspace{0.5 cm}
\caption{Nondimensional size distribution function $r_{k}(t)$ evaluated by solving the full 
model system of discrete equations (solid line) and the similarity solution (dashed line) at the 
nondimensional times 100, 200, 300, 400 and 500.}
\label{fig3}
\end{center}
\end{figure}

\begin{figure}
\begin{center}
\includegraphics[width=10cm]{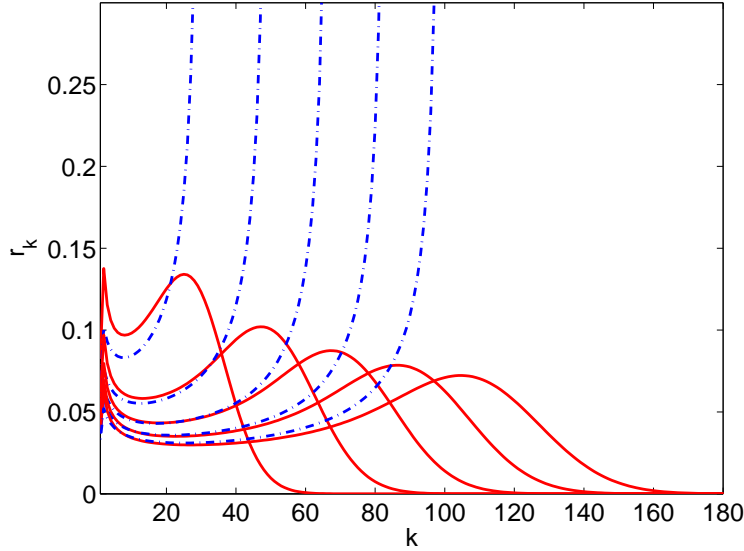}
\vspace{0.5 cm}
\caption{Same as Fig.~\ref{fig3} but now the dashed line is the similarity solution 
corrected by replacing $1+3s^{-1}-3\chi^{2/3}/2$ instead of $1-3\chi^{2/3}/2$ in 
(\ref{e48}).}
\label{fig2}
\end{center}
\end{figure}

The self-similar size distribution function has a singularity at its maximum size $\chi=
\chi_{0}$. Fig.~\ref{fig3} shows that this solution is a relatively poor approximation to 
the numerical solution of the full model. To improve it, we can  take into account the effects 
of discreteness by replacing $1+3s^{-1}-3\chi^{2/3}/2$ instead of $1-3\chi^{2/3}/2$ in 
(\ref{e48}). Then the singularity of the self-similar size distribution function moves closer 
to the local maximum of the exact numerical solution and the approximation improves, as 
shown in Figure \ref{fig2}. We observe that the singularity of the self-similar size 
distribution function occurs before the numerical size distribution function reaches its local 
maximum, and this effect becomes more noticeable as time increases. 

\subsection{Outer solution for the discrete problem}
The similarity solution of the continuum equations is a poor approximation to the numerical
solution of the discrete model. We need to correct it by including the effects of discreteness
and by approximating better the equation for the monomer concentration. The corrections
due to discreteness can be found by first solving the linear equations (\ref{e17}) exactly. 
Laplace transforming (\ref{e17}) and using $r_{k}(0)=0$, we obtain
\begin{eqnarray}
\hat{r}_{k}(\sigma) = {(k-1)^{1/3}\hat{r}_{k-1}\over\sigma + k^{1/3}}, \label{e25}
\end{eqnarray}
for $k>2$, from which it follows:
\begin{eqnarray}
&& k^{1/3}\,\hat{r}_{k}(\sigma) = 2\,\hat{c}(\sigma)\,\hat{R}_{k}(\sigma),
\label{e26}\\ 
&& \hat{R}_{k}(\sigma)\equiv \prod_{j=2}^k{1\over 1+\sigma\, j^{-1/3}}, 
\label{e27}\\ 
&& r_{k}(s) = 2 k^{-1/3}\int_{0}^s  R_{k}(s-s')\, c(s')\, ds'. \label{e28} 
\end{eqnarray}
These equations give the exact size distribution function in terms of $c(s)$. Inserting 
(\ref{e28}) in the definition of $M_{1/3}$ and substituting the result in (\ref{e21}), we 
obtain
\begin{eqnarray}
c\, {dc\over ds} + 4 c^2 + 2\, c\, \int_{0}^s  [\sum_{k=2}^\infty R_{k}(s-s')]\, 
c(s')\, ds' = 1. \label{e29} 
\end{eqnarray}
This integrodifferential equation for $c$ should be solved using the initial condition 
$c(0)=0$. Unfortunately, it is not too useful because the integral kernel is written as an
infinite series of the inverse Laplace transform of (\ref{e27}), which is rather unwieldly.

To find an approximate form of Eq.\ (\ref{e29}), we assume that $R_{k}(s)$
is peaked about its mean value:
\begin{eqnarray}
a(k)\equiv {\int_{0}^\infty s\, R_{k}(s)\, ds\over \int_{0}^\infty R_{k}(s)\, ds}
= -{\hat{R}'_{k}(0)\over \hat{R}_{k}(0)} = \sum_{j=2}^k j^{-1/3}. \label{e30}
\end{eqnarray}
In Appendix \ref{appA}, we show that 
\begin{eqnarray}
a(k)\sim {3\over 2}\, k^{2/3} - 3 + {1\over 2\,k^{1/3}}\quad (k\to\infty). 
\label{e31}
\end{eqnarray}
 We also show that the relative width of the peak in $R_{k}(s)$, $\sigma/a\sim 2 k^{-1/2}/
\sqrt{3}$ tends to 0 as $k\to\infty$. If we assume that $c(s)$ is a smooth function 
(and this is not always the case), then we may approximate the kernel in the convolution 
integral (\ref{e28}) by $R_{k}(s)\sim \delta\left(s-a(k)\right)$, so that the size
distribution function is given by (\ref{e42}) with $a(k)$ replaced by (\ref{e31}). 
Note that the first term in the right hand side of (\ref{e31})  coincides exactly with the 
result (\ref{e42a}) obtained by solving the continuum equations (\ref{e39}) and (\ref{e41}).
Equation (\ref{e29}) becomes
\begin{eqnarray}
c\, {dc\over ds} + 4 c^2 + 2c\, \sum_{k=2,\, a(k)<s} c(s-a(k)) = 1. \label{e36}
\end{eqnarray}
For large $s$, the position of the local maximum of the cluster size distribution is such that 
$a(k_{M})\sim 3 k_{M}^{2/3}/2 -3=s$. Then we can approximate the sum in (\ref{e36}) 
by an integral over $k$ from $k=0$ to $k=k_{M}(s)$. Changing variables in the integral, 
(\ref{e36}) becomes
\begin{eqnarray}
c\, {dc\over ds} + 4 c^2 + 2\,\sqrt{{2\over 3}}\, c\, \int_{0}^s
(s-s'+3)^{1/2}\, c(s')\, ds' = 1. \label{e37}
\end{eqnarray}

We would like to compare now the solution of (\ref{e42}), (\ref{e31}), (\ref{e37}) and 
(\ref{e20}) with zero initial conditions to the numerical solution of the exact discrete model. 
To this end, it is better to write all unknowns as functions of the time $t$. Our local theory is 
then
\begin{eqnarray}
&& r_{k}(t) = 2 k^{-1/3}\, c(s(t)-a(k))\, \theta(s(t)-a(k)), 
\nonumber\\
&& {dc\over dt} + 4 c^2 + 2\,\sqrt{{2\over 3}}\, c\, \int_{0}^{t} [s(t)-s(t')+3]^{1/2}
\, [c(s(t'))]^2\, dt' = 1, \label{e37-t}\\
&& {ds\over dt} = c, \nonumber\\
&& c(0)= s(0)= 0. \nonumber
\end{eqnarray}

\begin{figure}
\begin{center}
\includegraphics[width=10cm]{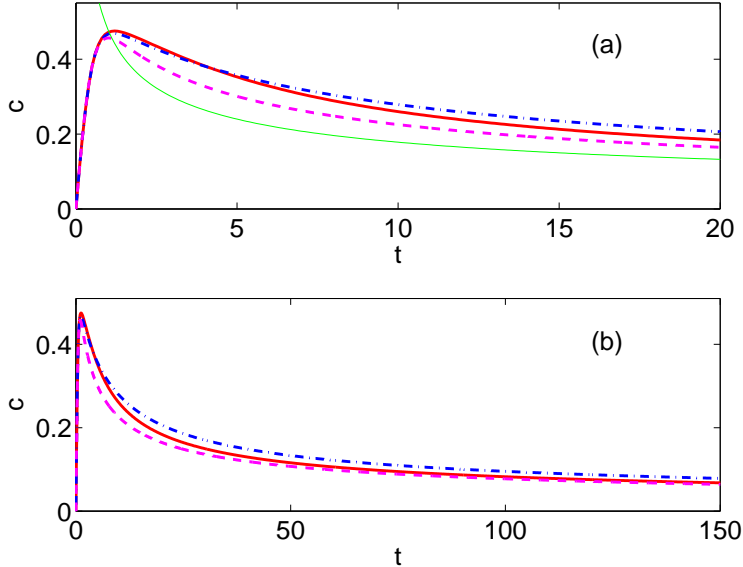}
\vspace{0.5 cm}
\caption{(a) Monomer concentration $c(t)$ evaluated using: (i) the numerical solution of 
the discrete equations of the model (solid line), (ii) the local theory (\ref{e37-t}) (dashed 
line), (iii) Schaldach and Wolfer's moment equations (dot-dashed line), and (iv) the 
self-similar solution having a vertical asymptote at $t=0$ (thin solid line). (b) Same as in (a) 
for a larger range of times. All variables are written in dimensionless units.}
\label{fig1}
\end{center}
\end{figure}

\begin{figure}
\begin{center}
\includegraphics[width=10cm]{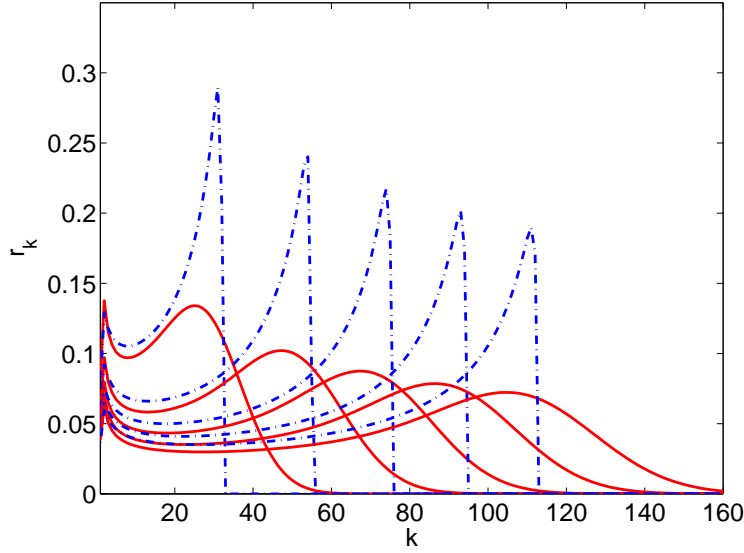}
\vspace{0.5 cm}
\caption{Dimensionless size distribution function $r_{k}(t)$ calculated using the numerical 
solution of the full model discrete equations (solid line) and the local theory (\ref{e37-t}) 
(dashed line) at the nondimensional times 100, 200, 300, 400 and 500.}
\label{fig5}
\end{center}
\end{figure}

Figure \ref{fig1} shows a comparison between the monomer concentration evaluated by
solving the full discrete model equations, the local theory (\ref{e37-t}), the similarity 
solution and the three moment equations used by Wolfer and coworkers (cf. Section 
\ref{sec:wolfer} and \cite{SW04}). In Fig.~\ref{fig1}(a), we observe that Wolfer et al's 
approximation is better for short times, but that the local theory given by (\ref{e37-t}) 
provides the best approximation as time goes to infinity, cf.\ Fig.~\ref{fig1}(b). Figure
\ref{fig5} shows the size distribution function calculated at different times by using the
local theory (\ref{e37-t}) (dashed lines) and the numerical solution of the full discrete model.
The local theory yields higher cluster densities than the exact solution (which is also the case 
with the self-similar solution in Fig.~\ref{fig2}), a much higher maximum
density but it predicts a maximum size which is very close to the local maximum of the real
distribution function at large sizes.  

\section{Leading edge of the size distribution function}
\label{front}
The previous local description of the size distribution function differs substantially from the
numerical solution of the model equations for large sizes. However, the maximum of the
numerical $r_{k}$ coincides with the peak of the approximate $r_{k}$ at $k=k_{M}(s)$, 
which is also its leading edge. To improve our asymptotic theory, we should insert a moving 
boundary layer there. How? We shall use the wave front theory which we introduced in our 
previous work on homogeneous nucleation \cite{NBC05}. Firstly, let us rewrite 
(\ref{e17}) and (\ref{e21}) as
\begin{eqnarray}
&& {d\sigma_{k}\over ds} = k^{1/3}\,(\sigma_{k-1}-\sigma_k),  \quad  k\geq 3,  
\label{f1}\\ 
&&\sigma_{k}= k^{1/3}r_{k},  \label{f2}\\
&& c\, {dc\over ds}+ 4\, c^2 + c\, \sum_{k=2}^\infty \sigma_{k} = 1.  \label{f3}
\end{eqnarray}

Secondly, let us use local coordinates about the inflection point of the wave front leading 
edge in $\sigma_{k}$, $k=K(s)$, for large $k$:
\begin{eqnarray}
\sigma_{k} = S(X,s), \quad X= k-K(s), \quad 1\ll X\ll K.    \label{f4}
\end{eqnarray}
Substitution of (\ref{f4}) in (\ref{f1}) yields
\begin{eqnarray}
{\partial S\over\partial s} - {\partial S\over\partial X}\, {dK\over ds} =
(K^{1/3} + {1\over 3} K^{-2/3} X+\ldots)\, \left(- {\partial S\over\partial X} +
{1\over 2}{\partial^2 S\over\partial X^2}+\ldots \right).    \nonumber
\end{eqnarray}
Provided
\begin{eqnarray}
{dK\over ds} = K^{1/3},  \label{f5}
\end{eqnarray}
the distinguished limit of the previous equation for $S$ gives
\begin{eqnarray}
{\partial S\over\partial s} + {1\over 3} K^{-2/3} X\, {\partial S\over\partial 
X} = {1\over 2} K^{1/3} {\partial^2 S\over\partial X^2}.    \label{f6}
\end{eqnarray}
A more detailed derivation using a book-keeping small parameter is included in Appendix 
\ref{appB}. Note that the inflection point $K(s)$ is between the position of the local 
maximum of the distribution function for large sizes and the maximum size, $k_{M}<K$. 
If we change variables from the adaptive time $s$ to the front location $K$, we find an 
equation in which $K$ scales as $X^2$. Thus it is convenient to consider $S$ as a function
of the new `time' $K$ and the similarity variable $\xi=X/\sqrt{K}$. Then,
\begin{eqnarray}
&& K\, {\partial S\over\partial K} - {\xi\over 6}\, {\partial S\over\partial\xi} 
= {1\over 2}\, {\partial^2 S\over\partial\xi^2},    \label{f7}\\
&& \xi = {k-K(s)\over\sqrt{K(s)}}. \label{f8}
\end{eqnarray}
Equation (\ref{f7}) should be solved with boundary condition $S(\infty,K)=0$ and an
appropriate matching condition as $\xi\to -\infty$. 

The solution of (\ref{f5}) is
\begin{eqnarray}
K(s)= \left({2s\over 3} + K_{0}^{2/3}\right)^{3/2},  \label{f9}
\end{eqnarray}
in which $K_{0}$ is an arbitrary positive constant to be selected later. The solution 
(\ref{e41}) with $a(k)\sim -3+3k^{2/3}/2$ yields the matching condition 
\begin{eqnarray}
\sigma_{k}= 2\, c\left(s+3-{3\over 2}\, k^{2/3}\right)_{+}\sim 2\, 
c\left(3-{3\over 2}\, K_{0}^{2/3} -\xi\,K^{1/6}\right)_{+},  \label{f10}
\end{eqnarray}
in the overlap region: $\sqrt{K}\ll (K-k)\ll K$, as $\xi\to -\infty$. In Appendix 
\ref{appC}, we show that the solution of (\ref{f7}) satisfying boundary and matching 
condition is 
\begin{eqnarray}
S(\xi,K) = {2\over\sqrt{6\pi (K^{1/3}-K_{0}^{1/3})}}\,
\int_{0}^{t} [c(t')]^2\, e^{-{\left[\xi K^{1/6}+ s(t')-3+{3\over 2}K_{0}^{2/3}
\right]^{2}\over 6\, (K^{1/3}-K_{0}^{1/3})}}\, dt'. \label{f11}
\end{eqnarray}

Clearly, the front solution (\ref{f11}) contributes to the moment $M_{1/3}$ in (\ref{e21})
for small times corresponding to $k$ in the overlap region. Since $S$ is matched to a variable
outer solution, it is convenient to pick a time $t_{p}(t)$ corresponding to $k$ in the overlap
region to split the time interval $(0,t)$ in (\ref{e37-t}) into two subintervals. For $0<t'<
t_{p}(t)$, we use the front approximation (\ref{f2}), (\ref{f4}) and (\ref{f11}) whereas
we use the outer approximation (\ref{e41}) with (\ref{e31}) for times in $(t_{p}(t),t)$.
The patching time solves
\begin{eqnarray}
s(t_{p}) = \xi_{p}\,\left[K(s(t))\right]^{1/6}+ 3- {3\over 2}\, 
K_{0}^{2/3},    \label{f12}
\end{eqnarray}
for $\xi_{p}$ in the overlap region. Since the width of the gaussian in (\ref{f11}) is
$\sqrt{6}$, we may choose $\xi_{p}\geq \sqrt{6}$. Up to the 
patching time, the contribution of the leading edge (\ref{f11}) to the moment $M_{1/3}$ in 
(\ref{e21}) is $\int_{-\infty}^\infty S(\xi,K)\, K^{1/2} d\xi = 2\, K^{1/3}
\int_{0}^{t_{p}} [c(t')]^2\, dt'$. Taking into account the approximation (\ref{e41}), we 
obtain 
\begin{eqnarray}
{dc\over dt}+ 4\, c^2 + 2\sqrt{{2\over 3}}\, c\, \int_{t_{p}(t)}^t [s(t)-s(t') + 
3]^{1/2} [c(t')]^2 dt'\nonumber\\
+ 2\, [K(s(t))]^{1/3}\, c\, \int_0^{t_{p}(t)} [c(t')]^2 dt' = 1.  \label{f13}
\end{eqnarray}

\begin{figure}
\begin{center}
\includegraphics[width=10cm]{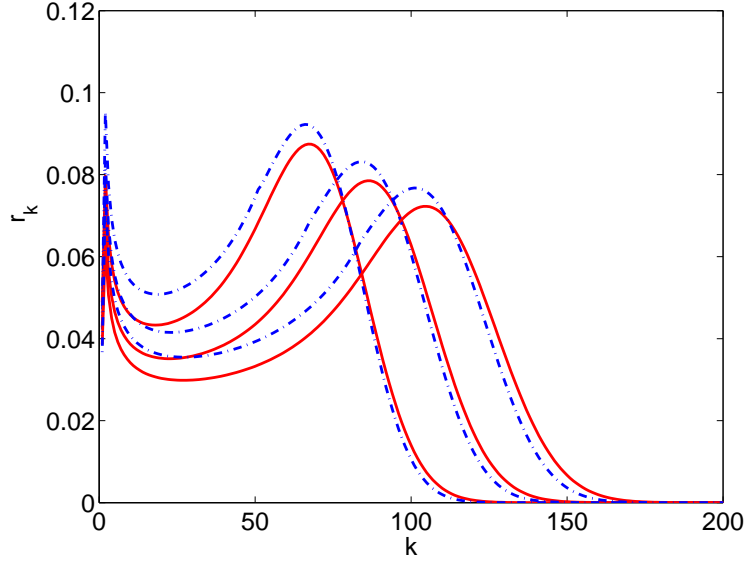}
\vspace{0.5 cm}
\caption{Nondimensional size distribution function $r_{k}(t)$ evaluated using the 
composite solution (\ref{f13}) - (\ref{f14}) (dashed line) and the numerical solution of 
the full model discrete equations (solid line) at the nondimensional times 300, 400 and 500.}
\label{fig4}
\end{center}
\end{figure}

Fig.~\ref{fig4} compares the numerical solution of the full discrete model equations to the 
composite solution:
\begin{eqnarray}
&& r_{k}= 2 k^{-1/3} c(s(t)-a(k))_{+}\theta(K-\xi_{p}\sqrt{K}-k) \nonumber\\
&&\quad +
{2\,\theta(k-K+\xi_{p}\sqrt{K})\over\sqrt{6\pi (K^{1/3}-K_{0}^{1/3})}}\,
\int_{0}^{t} [c(t')]^2\, e^{-{\left[\xi K^{1/6}+ s(t')-3+{3\over 2}K_{0}^{2/3}
\right]^{2}\over 6\, (K^{1/3}-K_{0}^{1/3})}}\, dt', \label{f14}
\end{eqnarray}
 plus (\ref{f13}) for the monomer concentration. We have used the numerical values 
 $K_{0}=0.5$ and $\xi_{p}=\sqrt{6}$ without looking for an optimal fit to the numerical 
 solution of the full model equations by varying $K_{0}$ and $\xi_{p}$. The agreement 
 between our theory and the numerical solution of the full model is much better than that 
 achieved by using only the outer solution. Although our effective small parameter is the 
 reciprocal of the wave front location (and $1/K\to 0$ as $t\to\infty$), the poor
 agreement between the outer solution and the numerical solution at large sizes precludes a
 better agreement between the composite solution and the numerical solution.
 
\section{Moment closure}
\label{sec:wolfer}
The composite solution is computationally costly if all we want to calculate is the monomer
concentration because we need to calculate $K(s(t))$, $t_{p(t)}$ and $c(s-a(k))$ at each 
instant of time. A different approximation involving only equations that
are local in time was used by Schaldach and Wolfer \cite{SW04} to approximate
the number densities of monomers, bubbles and helium. Note that Equations (\ref{e21}), 
(\ref{e23}) and (\ref{e24}) would form a closed system of equations if we could express
$M_{1/3}$ in terms of $M_{0}$ and $M_{1}$. Physically speaking, $M_{1/3}/M_{0}$ and 
$M_{1}/M_{0}$ have the meaning of average bubble radius and average helium density. If 
we impose
\begin{eqnarray}
{M_{1}\over M_{0}}\approx {4\pi\over 3}\, \left({M_{1/3}\over M_{0}}
\right)^3\Longrightarrow M_{1/3}\approx\left({3\over 4\pi}\right)^{1/3}\,
M_{1}^{1/3} M_{0}^{2/3}, \label{e59}
\end{eqnarray}
we obtain the following closed system of three differential equations \cite{SW04}
\begin{eqnarray}
&& {dM_{0}\over ds} = 2 c,  \label{e60}\\
&& {dM_{1}\over ds} = 4\, c + \left({3\over 4\pi}\right)^{1/3}\,
M_{1}^{1/3} M_{0}^{2/3}, \label{e61}\\
&& c\, {dc\over ds} + 4 c^2 + \left({3\over 4\pi}\right)^{1/3}\,
c\, M_{1}^{1/3} M_{0}^{2/3} = 1, \label{e62}
\end{eqnarray}
plus Eq.\ (\ref{e20}) relating $s$ and the time. For long times, these equations possess the 
same scaling symmetry as the continuum equations and their solutions tend to a similarity 
solution for long times. A different closure assumption can be obtained by combining the idea
of preserving the scaling symmetry with a closure assumption related to (\ref{e59}), as
explained in Appendix \ref{appD}. The resulting moment equations have solutions which 
become self-similar for appropriate intervals of time, but these solutions approximate the
numerical solution of the kinetic equation less well than those of (\ref{e60}) - (\ref{e62}).  

\begin{figure}
\begin{center}
\includegraphics[width=10cm]{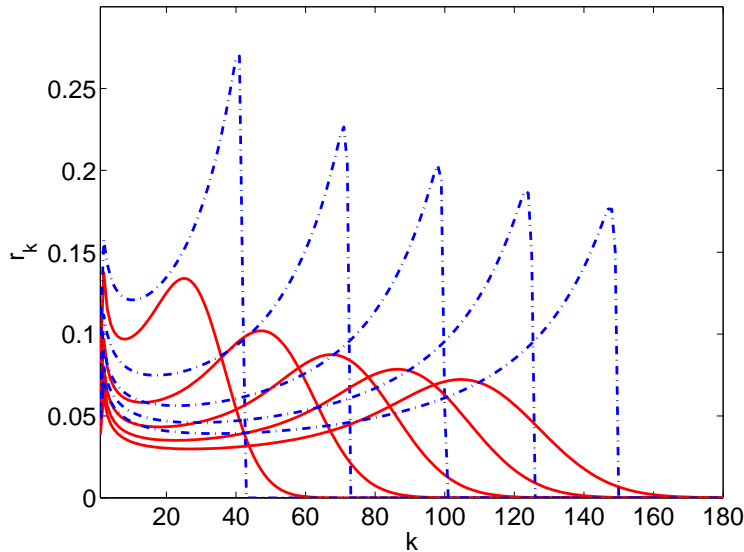}
\vspace{0.5 cm}
\caption{Nondimensional size distribution function $r_{k}(t)$ of (\ref{e42}) and 
(\ref{e31}) with monomer concentration given by the moment equations (\ref{e60}) - 
(\ref{e62}) (dashed line) and the numerical solution of the full model discrete equations 
(solid line). Times are as in Fig.~\ref{fig1}.}
\label{fig7}
\end{center}
\end{figure}

Figure \ref{fig7} shows a comparison between the size distribution function (\ref{e42}) 
and (\ref{e31}) calculated with the monomer concentration resulting from the moment 
equations (\ref{e60}) - (\ref{e62}) and the numerical solution of the full model. We 
observe that the maximum at large sizes is predicted to occur at much larger sizes. Thus the 
moment equations provide a worse prediction than the outer solution (\ref{e37-t}), and, of
course, worse than the composite solution (\ref{f13}) - (\ref{f14}). 

\section{Conclusions}
\label{discusion}
In Section \ref{similarity}, we have found the composite solution (\ref{f13}) - 
(\ref{f14}) to the discrete kinetic equations. The outer approximation is a solution of the
continuum limit of the discrete equations corrected by the effects of discreteness. The inner 
approximation follows from a wave front expansion previously introduced for transient
homogeneous nucleation \cite{NBC05}. A similar equation for the wave front profile was 
obtained by King and Wattis in the case of the Becker-D\"oring equations with rate constants 
having a power law dependence on cluster size \cite{WKi02}. While in the case of nucleation 
the outer approximation was a simple constant profile in the wake of the wave front, in the 
present case of growing helium bubbles, the outer approximation is a time and size dependent 
function depending on the monomer concentration. The monomer concentration satisfies
an integrodifferential equation comprising contributions of both the outer and the inner
solutions. As time increases, the outer solution tries to achieve a self-similar form in the 
variable $\chi=k/s^{3/2}$ corrected by discreteness effects, as suggested by Fig.~\ref{fig2}.
The inner solution is determined by the position of the wave front $K(s(t))$ and by the 
similarity variable $\xi=(k-K)/\sqrt{K}$, which is different from $\chi$. Inner and outer
solution have an overlap domain $1\ll\sqrt{K}\ll (K-k)\ll K$ and are patched at a 
point thereof to yield the composite solution. 

We have also compared our approximations to Schaldach and Wolfer's moment closure of 
Section \ref{sec:wolfer} supplemented with the formulas (\ref{e42}) and (\ref{e31}) for 
the size distribution function. Fig.~\ref{fig1} shows that the similarity solution for $c(t)$ is 
consistently above the numerical solution of the model, while the moment closure solution is 
consistently below. Then the maximum of the size distribution function approximated using 
the similarity solution (resp.\ the moment closure theory) occurs before (resp.\ after) the 
real maximum. In contrast with these approximations, the outer solution (\ref{e37-t}) yields 
a better match to the monomer concentration for large times and to the location of the 
maximum of the size distribution function. Corrected with the boundary layer formula, the 
composite solution (\ref{f13}) - (\ref{f14}) gives the best match to the numerical solution 
of all the approximations described in this paper. 

\section*{Acknowledgements}
This work has been supported by the Spanish MEC grants MAT2005-05730-C02-01 and
MAT2005-05730-C02-02, and by the US NSF grant 0515616. 

\appendix
\section{Approximating the sums $\sum_{j=2}^k j^{-\alpha}$.}
\label{appA}
For $0<\alpha<1$ and $k\to\infty$, these expressions are Riemann sums whose
$j$th term equals the area under a rectangle of height $(j-1)^{-\alpha}$ and basis 1.
These Riemann sums can be approximated by the integral minus the sum of the areas
of triangles of height $[(j-1)^\alpha-j^\alpha]$ and basis 1. We thus have
$$
\sum_{j=2}^k j^{-\alpha}=  \sum_{j=1}^k j^{-\alpha} -1 \approx k^{1-\alpha}
\int_{1/k}^1 {dx\over x^\alpha} + {1\over 2}\, \sum_{j=2}^k [j^{-\alpha} - 
(j-1)^{-\alpha}] - 1, $$
and therefore
$$
\sum_{j=2}^k j^{-\alpha}\approx {k^{1-\alpha}\over 1-\alpha} - {5-3\alpha\over
2\, (1-\alpha)} + {1\over 2 k^\alpha}.
$$
The function $a(k)$ is given by the sum with exponent 1/3, which yields (\ref{e31}). 
Similarly, the half-width of the peak is given by $\sigma(k)$, 
\begin{eqnarray}
\sigma(k)^2 &\equiv& \int_{0}^\infty [s-a(k)]^2\, R_{k}(s)\, ds =
\hat{R}''(0)-[\hat{R}'(0)]^2 = \sum_{j=2}^k j^{-2/3}\nonumber\\ 
& \sim & 3 k^{1/3} - {9\over 2} + {1\over 2\,k^{2/3}}\quad (k\to\infty). 
\nonumber
\end{eqnarray}
As explained in the text, the relative width of the peak of $R_{k}(s)$, $\sigma/a\sim 2 
k^{-1/2}/\sqrt{3}$ tends to 0 as $k\to\infty$.

\section{Derivation of the equation for the wave front profile using a book-keeping small
parameter.}
\label{appB}
Let us insert 
\begin{eqnarray}
\sigma_{k} = S(X^*,s^*), \quad X^* = \epsilon^\gamma\,\left(k-{K^*(s^*)\over\epsilon}
\right), \quad s^* = \epsilon^{2/3} s, \quad \epsilon\to 0+,    \nonumber
\end{eqnarray}
in (\ref{f1}) instead of (\ref{f4}). The small parameter $\epsilon$ represents location of
the wave front at a typical large size. We find
\begin{eqnarray}
\epsilon^{2/3}{\partial S\over\partial s^*} - \epsilon^{\gamma-1/3} {\partial 
S\over\partial X^*}\, {dK^*\over ds^*} = \left({(K^*)^{1/3}\over\epsilon^{1/3}}
+ {\epsilon^{2/3-\gamma}\over 3} (K^*)^{-2/3} 
X^* +\ldots\right)\nonumber\\
\times \left(- \epsilon^\gamma {\partial S\over\partial X^*} + {\epsilon^{2
\gamma}\over 2}{\partial^2 S\over\partial (X^*)^2}+\ldots \right),   \nonumber
\end{eqnarray}
from which
\begin{eqnarray}
\epsilon^{2/3}\left( {\partial S\over\partial s^*} + {X^*\over 3 (K^*)^{2/3}} 
{\partial S\over\partial X^*}\right) + \epsilon^{\gamma-1/3} {\partial S\over
\partial X^*}\, \left((K^*)^{1/3} - {dK^*\over ds^*}\right) = \nonumber\\
{\epsilon^{2\gamma-1/3}(K^*)^{1/3}\over 2}{\partial^2 S\over\partial (X^*)^2} 
+\ldots   \nonumber
\end{eqnarray}
Provided
\begin{eqnarray}
 {dK^*\over ds^*} =(K^*)^{1/3},  \nonumber
\end{eqnarray}
and $\gamma=1/2$, we find the equation
\begin{eqnarray}
{\partial S\over\partial s^*} + {X^*\over 3(K^*)^{2/3}} {\partial S\over
\partial X^*} = {(K^*)^{1/3}\over 2}{\partial^2 S\over\partial (X^*)^2} ,
\nonumber
\end{eqnarray}
in the limit as $\epsilon\to 0+$. The previous two equations are (\ref{f5}) and 
(\ref{f6}) once we revert to the variables $s$ and $K$.

\section{Solution of the equation for the leading front}
\label{appC}
Defining $J=- \partial S/\partial\xi$, we find the following equation for $J$:
$$ K\, {\partial J\over\partial K} - {1\over 6}\, {\partial (\xi J)\over\partial
\xi} = {1\over 2}\, {\partial^2J\over\partial\xi^2}.
$$
The Fourier transform of $J(\xi,\cdot)$ satisfies the hyperbolic equation
$$ K\, {\partial\hat{J}\over\partial K} + {\eta\over 6}\, {\partial\hat{J}
\over\partial\eta} = -{\eta^2\over 2}\, \hat{J},
$$
which is readily solved by the method of characteristics in terms of an arbitrary initial 
condition $\hat{J}_{0}(\eta_{0})$. Inverting the Fourier transform and going back to 
the function $S$, we obtain
\begin{eqnarray}
S(\xi,K) = \mathcal{S}(K) + {1\over\sqrt{6\pi [1-(K_{0}/K)^{1/3}]}}\,
\int_{-\infty}^\infty S_{0}(\xi_{0})\, \exp\left[-{(\xi-\xi_{0})^{2}\over 6\,
[1-(K_{0}/K)^{1/3}]}\right]\, d\xi_{0}, \nonumber
\end{eqnarray}
in which $\mathcal{S}(K)$ and $S_{0}(\xi_{0})$ are both arbitrary. As $\xi\to -\infty$,
the integral can be approximated by the Laplace method with the result $S(\xi,K)\sim 
\mathcal{S}(K) + S_{0}(\xi)$. The matching condition (\ref{f10}) gives $S_{0}\sim
2\, c(3-K^{1/6}\xi-3K_{0}^{2/3}/2)- \mathcal{S}(K)$, thereby yielding
\begin{eqnarray}
S(\xi,K) = {2\over\sqrt{6\pi [1-(K_{0}/K)^{1/3}]}}\, \int
c\left(K^{1/6}\xi_{1}+3-{3K_{0}^{2/3}\over 2}\right)\, e^{-{(\xi+\xi_{1})^{2}
\over 6\, [1-(K_{0}/K)^{1/3}]}}\, d\xi_{1}. \nonumber
\end{eqnarray}
Changing variables from $\xi_{1}$ to $s'=K^{1/6}\xi_{1}+3-3K_{0}^{2/3}/2$, we
find
\begin{eqnarray}
S(\xi,K) = {2\over\sqrt{6\pi  (K^{1/3}-K_{0}^{1/3})}}\, \int_{0}^s c(s')\, 
e^{-{\left(K^{1/6}\xi-3+{3K_{0}^{2/3}\over 2}+s'\right)^{2}\over 6\, (K^{1/3}-
K_{0}^{1/3})}}\, ds'. \label{c1}
\end{eqnarray}
The ends of the integration interval in this expression are set to 0 and $s$ because those are the
extremes of the interval over which the monomer concentration exists. Changing variables 
in this formula from $s'$ to the time $t'$, we obtain (\ref{f11}). 

The width of the Gaussian in (\ref{c1}) is $\sqrt{6}$. Thus a typical point in the overlap 
region is $k_{p}= K - \xi_{p}\sqrt{K}$ with $\xi_{p}\geq\sqrt{6}$. The corresponding 
adaptive time is
$$s_{p} = s+3- {3\over 2} k_{p}^{2/3}\sim \xi_{p} [K(s)]^{1/6} + 3- {3\over 2}\, 
K_{0}^{2/3},
$$
and the corresponding time is given by (\ref{f12}).

\section{Moment equations following from a closure assumption preserving scaling
symmetry}
\label{appD}
Let us assume that the size distribution function has the scaling form
\begin{eqnarray}
&& r_{k}(s) = {M_{0}(s)^2 l_{1}\over M_{1}(s) l_{0}^2}\, \tilde{r}(x), \label{e63}\\
&& x= {M_{0}(s) l_{1}\over M_{1}(s) l_{0}}\, k,Ê\label{e64}\\
&& l_{\mu}= \int_{0}^\infty x^\mu \tilde{r}(x)\, dx.   \label{e65}
\end{eqnarray}
The definition (\ref{e22}) of the moments together with (\ref{e63}) - (\ref{e65}) 
imply
\begin{eqnarray}
M_{\mu} = {l_{\mu}\over  l_{1}^\mu\, l_{0}^{1-\mu}}\, M_{1}^\mu M_{0}^{1-\mu}.  
\label{e66}
\end{eqnarray}
We can use (\ref{e66}) for $\mu=1/3$ to close the system of equations (\ref{e21}), 
(\ref{e23}) and (\ref{e24}), with the result
\begin{eqnarray}
&& {dM_{0}\over ds} = 2 c,  \label{e67}\\
&& {dM_{1}\over ds} = 4\, c + \lambda\, M_{1}^{1/3} M_{0}^{2/3}, \label{e68}\\
&& c\, {dc\over ds} + 4 c^2 + \lambda\, c\, M_{1}^{1/3} M_{0}^{2/3} = 1, 
\label{e69}\\
&& {ds\over dt} = c, \label{e70}\\
&& \lambda = {l_{1/3}\over  l_{1}^{1/3}\, l_{0}^{2/3}}. \label{e71}
\end{eqnarray}
These equations become (\ref{e60}) - (\ref{e62}) if $\lambda = (4\pi/3)^{-1/3}$. To 
make them compatible with the previously found similarity solution, we consider the reduced
system given by (\ref{e67}) and the approximate equations
\begin{eqnarray}
{dM_{1}\over ds} = \lambda\, M_{1}^{1/3} M_{0}^{2/3} = {1\over c}, \label{e72}
\end{eqnarray}
which have the same scaling symmetry as in the case of the continuum limit. Thus they have
the similarity solution 
\begin{eqnarray}
c = C_{1} s^{-3/4}, \label{e73}
\end{eqnarray}
for a certain constant $C_{1}$. From (\ref{e67}) and (\ref{e72}) (written as $dM_{1}/ds
= 1/c$), we obtain
\begin{eqnarray}
 M_{0} = 8\, C_{1} s^{1/4}, \quad M_{1}= {4\, s^{7/4}\over 7 C_{1}}, \label{e74}
\end{eqnarray}
whereby (\ref{e72}) yields 
\begin{eqnarray}
C_{1} = {7^{1/4}\over 4}\, \lambda^{-3/4}. \label{e75}
\end{eqnarray}

We shall now use Eq.\ (\ref{e48}) for $r_{k}$ and (\ref{e63}) and (\ref{e73}) -
(\ref{e75}) to find $\tilde{r}$. Straightforward but lengthy calculations yield
\begin{eqnarray}
&& \tilde{r}(x) = { l_{0}\over 6\, x_{M}}\,\left({x\over x_{M}}\right)^{-1/3} 
\left[1 - \left({x\over x_{M}}\right)^{2/3}\right]_{+}^{-3/4}, \label{e76}\\
&& x_{M}= \left({7\, l_{1}\over 6\, l_{1/3}}\right)^{3/2}. \label{e77}
\end{eqnarray}
With this reduced size distribution function we can check that (\ref{e65}) becomes $l_{0}=
l_{0}$ for $\mu=0$ whereas it becomes
\begin{eqnarray}
&& l_{1}= \sqrt{{7\, l_{1}^3\over 3\pi l_{1/3}^3}}\, {l_{0}\, \Gamma(1/4)^2\over 
12}, \label{e78}\\
&& l_{1/3} = \sqrt{{ 3\, l_{1/3}\over 7\pi l_{1}}}\, {l_{0}\, \Gamma(1/4)^2\over 6}, 
\label{e79}
\end{eqnarray}
for $\mu=1$ and 1/3, respectively. These two last equations and (\ref{e77}) imply that
\begin{eqnarray}
l_{1/3} = {7\over 6}\, l_{1}, \quad x_{M}=1.   \label{e80}
\end{eqnarray}
Thus (\ref{e80}) are required for the reduced size distribution function to be consistent with
the definitions of the $l_{\mu}$. Using (\ref{e78}) - (\ref{e80}) in (\ref{e71}) and 
(\ref{e75}), it is possible to show that $C_{1}=R_{0}/2$ given by (\ref{e51}) and we
get the same similarity solution as before.

Had we used Schaldach and Wolfer's closure $\lambda = (4\pi/3)^{-1/3}$, we would
have obtained the following similarity solution to the reduced system (\ref{e67}) and
(\ref{e72}):
\begin{eqnarray}
c = C_{2} s^{-3/4},  \quad M_{0} = 4\, C_{2} s^{1/4}, \quad 
M_{1}= {4\over 7\, C_{2}}\, s^{7/4},\quad C_{2}= {1\over 2}\left({7\pi\over 3}
\right)^{1/4}.  \label{e81}
\end{eqnarray}
If we employ time instead of the variable $s$, (\ref{e81}) becomes 
\begin{eqnarray}
c = \left({4\pi\over 3}\right)^{1/7} 7^{-2/7} t^{-3/7},  \quad 
M_{0} = \left({4\pi\over 3}\right)^{2/7} 7^{3/7} t^{1/7}, \quad 
M_{1}= t.  \label{e82}
\end{eqnarray}

\end{document}